\documentclass[acmsmall]{acmart}
\makeatletter
\def\@ACM@checkaffil{
    \if@ACM@instpresent\else
    \ClassWarningNoLine{\@classname}{No institution present for an affiliation}%
    \fi
    \if@ACM@citypresent\else
    \ClassWarningNoLine{\@classname}{No city present for an affiliation}%
    \fi
    \if@ACM@countrypresent\else
        \ClassWarningNoLine{\@classname}{No country present for an affiliation}%
    \fi
}
\makeatother

\usepackage{xcolor}
\usepackage{amsmath}
\usepackage{xcoffins,calc}
\usepackage[most]{tcolorbox}

\newtcolorbox{mybox}{
enhanced, boxrule=0pt,frame hidden, borderline west={4pt}{0pt}{green!75!black}, colback=green!10!white, sharp corners
}

\AtBeginDocument{%
  \providecommand\BibTeX{{%
    \normalfont B\kern-0.5em{\scshape i\kern-0.25em b}\kern-0.8em\TeX}}}

\setcopyright{acmcopyright}
\copyrightyear{2024}
\setcopyright{rightsretained}
\acmJournal{PACMHCI}
\acmYear{2024} \acmVolume{8} \acmNumber{CSCW2} \acmArticle{508} \acmMonth{11}\acmDOI{10.1145/3687047}

\acmConference{CSCW 2024}{November 9-13, 2024}{Costa Rica}
\acmPrice{15.00}
\acmISBN{978-1-4503-XXXX-X/18/06}

\acmSubmissionID{V8cscw508}

\begin{document}

\title{Paradoxes of Openness and Trans Experiences in Open Source Software}

%


\author{Hana Frluckaj}
\affiliation{%
    \institution{University of Texas at Austin},
    \institution{Carnegie Mellon University}}

\author{James Howison}
\affiliation{%
    \institution{University of Texas at Austin}}
 
\author{Laura Dabbish}
\affiliation{%
  \institution{Carnegie Mellon University}}

\author{Nikki Stevens}
\affiliation{%
  \institution{Massachusetts Institute of Technology}
}

\renewcommand{\shortauthors}{Frluckaj, Howison, Dabbish, and Stevens}

\begin{abstract}
  In recent years, concerns have increased over the lack of contributor diversity in open source software (OSS), despite its status as a paragon of open collaboration. OSS is an important form of digital infrastructure and part of a career path for many developers. While there exists a growing body of literature on cisgender women’s under-representation in OSS, the experiences of contributors from other marginalized groups are comparatively absent from the literature. Such is the case for trans contributors, a historically influential group in OSS. In this study, we interviewed 21 trans participants to understand and represent their experiences in the OSS literature. From their experiences, we theorize two related paradoxes of openness in OSS: the paradox of openness and display and the paradox of openness and governance. In an increasingly violent world for trans people, we draw on our theorizing to build recommendations for more inclusive and safer OSS projects for contributors.
\end{abstract}

\ccsdesc[500]{Human-centered computing~Empirical studies in collaborative and social computing}

\keywords{open source software, gender, diversity, inclusion, open collaboration}

\received{14 July 2023}
\received[revised]{16 January 2024}
\received[accepted]{30 May 2024}

\maketitle

\section{Introduction}
Open source software (OSS) represents a critical form of digital infrastructure, collaborative learning environment, and a potential technical career path for many developers. Open source is also a successful example and an inspiration for open collaboration more broadly~\cite{levine_open_2014}. Yet both open source---and open collaboration in general---is surprisingly lacking in diversity, with poor representation of non-male and non-white participants~\cite{nadri_relationship_2022,shameer_relationship_2023,trinkenreich_womens_2022,frluckaj_gender_2022}. Indeed these rates of participation appear to be lower even than those found in the software industry more broadly~\cite{trinkenreich_womens_2022,bureau_of_labor_statistics_employed_2022}. Despite the promise of openness---online work where anyone can, in theory, participate---the reality is problematic both in terms of overall participation, dropout rates, and quality of experience~\cite{dunbar-hester_hacking_2019, dahlberg_cyber-libertarianism_2010}.

Based on a qualitative study of trans participants in OSS (here we use “trans” to describe anyone whose lived, desired, or experienced gender does not align with the gender assigned to them at birth~\cite{stryker_transgender_2008}), we develop theory to help understand the relationship between openness and diverse participation. Although there exists an established literature on women’s lower participation levels in OSS~\cite{dunbar-hester_hacking_2019, nafus_patches_2012}, it does not specify what portion of study participants are cis or trans women. Indeed, the experiences of trans contributors in OSS projects has rarely been mentioned, thus obfuscating the true representation of trans people in OSS.

The lack of focus from academic researchers on trans OSS participation is striking, given that trans contributors have played a vital role in both computer science~\cite{olyslager_calculation_2007, hiltzik_through_2000} and OSS over many years. Examples include Coraline Ada Ehmke's creation of the Contributor's Covenant~\cite{osland_how_2018}, Sage Sharp’s efforts to improve communication and inclusivity in Linux ~\cite{gold_torvalds_2012}, and Audrey Tang’s translation of OSS books to Chinese~\cite{li_taiwancnetcom_2006}. Despite the bigotry, harassment, and backlash trans contributors have faced in online and offline spaces~\cite{kidd_gamergate_2016,massanari_gamergate_2017,schoenebeck_drawing_2021}, their participation has been formative; these experiences reinforce the importance of learning from stories of trans participants to address discrimination and exclusion within open collaboration. The values, norms, and practices of a community are more salient to those on the margins, thus understanding how OSS communities engage with trans contributors is an indicative measure of project inclusivity and solidarity. We believe it is important to highlight the presence, contributions, and involvement of trans contributors in OSS, especially given the current climate of violence and political backlash against trans people~\cite{robinson_lavender_2020, mcclearen_if_2022}. 

Addressing the shortcomings of open collaboration in fostering nurturing environments is crucial. First, justice demands it; participation should not be a privilege reserved only for the majority. Second, diversity creates greater effectiveness and performance in teams~\cite{geiger_summary_2017}, better collaboration~\cite{shameer_relationship_2023}, higher earnings in organizations~\cite{mendez_open_2018}, with higher productivity in OSS~\cite{ford_paradise_2016,vasilescu_gender_2015}. Third, for individuals, participation is not only an important source of self-expression~\cite{dunbar-hester_hacking_2019} and collaborative learning, but also supports skill building and career opportunities by circumventing institutional hiring processes and gatekeeping~\cite{sharma_motivation-hygiene_2022,germonprez_theory_2017,marlow_activity_2013}. 
Finally, wider society benefits through the enhancement of individual opportunities, but also because OSS tools impact the world building infrastructures, algorithms, and applications used widely and daily~\cite{dunbar-hester_hacking_2019}. 
As OSS continues to grow~\cite{GitHub_state_2021} and to inspire open collaboration more broadly, it is crucial to ensure this growth supports diverse communities and resists societal inequities. This study therefore aims to add empirical evidence of the experiences of the trans community to the existing OSS literature, center the experiences of trans contributors, theorize our findings in the context of open collaboration, and therefore identify future directions for OSS inclusivity and authentic participation. 

\begin{mybox}
\textbf{Research Questions:} 
\begin{enumerate}
    \item What did trans participants highlight about their experiences in OSS?
    \item How did our participants manage their online identities in OSS?
\end{enumerate}
\end{mybox}

\section{Related Work}
\subsection{Queer HCI}
Queer HCI, or scholarship about LGBTQIA+ people in HCI, has grown steadily in recent years. "Queer" is used by HCI researchers to describe those who are not cisgender and/or heterosexual, however, it can be a difficult as an umbrella term, i.e. some trans people do not identify as queer~\cite{halberstam_queer_2011}. While queer HCI has previously centered gay men~\cite{spiel_queering_2019}, it has recently expanded to better encompass and center the increasingly diverse and evolving queer community, e.g. queer youth~\cite{homan_social_2014, badillo-urquiola_conducting_2021}. 

There exists a significant body of work on queer online identity management~\cite{odom_understanding_2015, farnham_faceted_2011, zhang_separate_2022}, including how being trans and visible online in everyday social media usage can be a form of political advocacy and activism~\cite{lerner_privacy_2020}. Indeed, recent studies on self-presentation ``advocate for supporting selective visibility in design
because the assumed isomorphism between one-account and oneself breaks down for those with heightened self-presentation needs''~\cite{taylor_cruising_2024}.

Queer HCI has tended to study broad LGTBQIA+ groups; researchers have called for future work to study more specific groups~\cite{taylor_cruising_2024} and intersectionality~\cite{walker_more_2020} and to decenter technology~\cite{taylor_cruising_2024}. Additionally, while many studies examine queer participation in social media~\cite{zhang_modeling_2015}, i.e. fear of being outed online~\cite{wyche_facebook_2013, devito_too_2018}, open collaboration and peer production environments such as OSS have yet to be examined. As noted above, we hope our paper creates opportunities for the OSS literature in covering experiences outside of men and (cis) women’s experiences, and we see opportunities to advance the Queer HCI literature through engagement with open collaboration and peer production contexts such as OSS. By providing empirical evidence of the experiences of trans OSS contributors we hope to bridge these two literatures.

\subsection{Queer Identity and Perceived Risks Online}
Gender is a highly complex and personal concept regarding one’s internal sense of themselves as a person, with many gender identities existing beyond the traditional binary dichotomy (i.e., male or female). The dissonance between a person’s gender and the gender assigned to them at birth can lead to negative effects, including distress, depression, and gender dysphoria~\cite{davy_what_2018}, that can be alleviated through transitioning. Those who transition aim to “better align their appearance, bodies, and/or legal documentation with their gender”~\cite{chuanromanee_designing_2022}. When transitioning, trans people often reject the name assigned to them at birth, also known as their “deadname,” in favor of a name of their choosing. Generally, to refer to someone by their deadname is a faux pas at best~\cite{sinclair-palm_its_2022}, and active discrimination or violence at worst~\cite{earle_deadnames_2022}. Digital footprints have varying levels of mutability, which is of particular importance when trans people change their names \cite{spiel_why_2021}. Reclaiming past work under one’s deadname, if even possible, places further strain on trans people’s time, energy, and resources. In fact, many academics and researchers have bemoaned the glacial adaptation to name changes and the resulting trans-exclusive inequity in publishing records~\cite{gaskins_visible_2021,hunt_impact_2020,tanenbaum_publishers_2020}. It is unclear how trans software developers and their work in OSS are impacted by these changes.

\subsubsection{Queer Identity Representation and Behavior Online}
Trans people often turn to online spaces to express themselves, to connect with other trans people, and to find resources and social support that are otherwise difficult to find offline~\cite{craig_you_2014}, especially in a world where gender-based violence against trans people continues to rise~\cite{wirtz_gender-based_2020}. Many have concerns about being accidentally outed while participating online, putting them at risk for harassment, trolling, and even physical harm~\cite{litt_imagined_2016, lucero_safe_2017, pinter_entering_2021}. 
As a result, trans people are often mindful of potential perceived audiences and tailor their privacy settings, profile visibility, and online behaviors~\cite{litt_knock_2012, scheuerman_safe_2018, duguay_he_2016, zhao_social_2016}, all parallels to women’s behavior in OSS~\cite{frluckaj_gender_2022}. It is important to note, however, that these documented experiences are racialized: “trans” and “women” in academic literature are nearly always white-coded~\cite{snorton_black_2017}. 

Zhao et al. examine how non-queer users manage multiple social media sites representations, indicating the ability to partition audiences and distribute content at the choosing of the creator~\cite{zhao_social_2016}. On the other hand, Duguay observed LGBTQ+ identity disclosure on social media sites that disclosure practices were influenced by pre-existing social conditions (e.g., would their coming out be accepted by their parents) and the participant’s perception of their information privacy~\cite{duguay_he_2016}. LGBTQ+ users were also influenced by their imagined audience for each social media sites and fine-tuned their profiles and accounts accordingly~\cite{litt_knock_2012, litt_imagined_2016}. Another work found that many LGBTQ+ individuals maintained a gap between their actual social identity, shared within a trusted community, and a virtual sanitized social identity, to be visible with for who might react negatively to their actual identity ~\cite{goffman_stigma_2009}. DeVito et al. found that these gaps varied amongst different social media sites based on the perceived audience of each site (e.g., certain content not posted because it was “too gay for Facebook”)~\cite{devito_too_2018}. Work by DeVito and Haimson indicate the importance of safety online for trans people as they become increasingly comfortable with representing their identity~\cite{haimson_designing_2020, devito_too_2018}.

\section{Methods}
Our goal in this study was to understand the experiences of trans people in OSS and any impact of gender identity on OSS career trajectory. To better investigate these experiences, we employed a qualitative methods approach to gather information, details, and insights from their involvement and participation. We interviewed 21 trans contributors from a variety of technical backgrounds, project types, and lived experiences. This study has received IRB approval.  

\subsection{Sampling and Recruitment Procedure}
Our sample began with a convenience sampling approach from prior connections with trans OSS contributors. We reached out to these contributors to see if they would be willing to participate in this study and if they knew any other contributors to contact. Thus began a snowballing approach that enabled us to connect with additional study participants. 

To supplement our snowball approach, we searched through GitHub platform’s user list for profiles with the words “trans” or “non-binary” in the profiles, and also searched for public profiles that had pronouns with some variation of “they/them” (e.g., they/she, it/they). While we recognize that pronoun choice is not a holistic reflection of gender identity, we relied on these pronouns as a way to identify individuals on GitHub that could be perceived as trans by other contributors. Utilizing pronouns helped us more readily locate individuals who were likely to be trans. The first author uses she/they pronouns, which helped the author convey allyship and connect with potential participants while recruiting. We recognize that this method might have inadvertently exposed individuals to scrutiny from academic researchers, a group that has exploited trans participants in the past~\cite{logie_we_2012}. We were mindful of the potential harm of pronoun recognition and the likelihood for misclassification in our approach~\cite{hamidi_gender_2018}. 

In total, we contacted 56 participants, of which 25 responded and 21 were interviewed (4 were unable to meet due to scheduling conflicts). Of our 21 participants, 4 were previous connections and 4 were recruited through snowballing, and 13 participants were found through GitHub’s or Twitter's user search function. Of the 8 that were not found through the search function, 6 were not publicly out at the time of our interview.

To accurately reflect the gender identity of each participant and follow best-practices for gender inclusivity in research as noted by~\cite{scheuerman_hci_2020, taylor_cruising_2024}, we asked individuals to self-identify their gender in an open response question during the interviews.

Eligibility criteria included being over 18, actively contributing to OSS in the past year, and a minimum of 10 public contributions during that time (the exception was P14, who had 4 contributions and was burnt out and taking a break from OSS participation). We used GitHub as a recruitment site due to its dual nature as a professional and personal/hobbyist environment. All contacts were through email, Twitter, or LinkedIn, and explicitly stated the study's purpose, ensured anonymity, and clarified the voluntary nature of participation. No compensation was provided to participants.

Demographic details for our participants are listed in Table 1 (Appendix A). Questions on their gender identities, racial and ethnic identities, and national origin were open response. 14 of our 21 participants were publicly out at the time of our interview. 13 out of 21 were Non-Hispanic and White, and only 5 of our participants did not originate from the US.

\subsection{Interviews}
The first author conducted semi-structured interviews, a type of flexible interviewing where interviewers can adjust questions as they see fit during their discussion with participants. This allowed for ample focus on the research questions we initially sought to answer while also leaving room for non-obvious but relevant content. On average, the interviews ran about 55 minutes (durations ranged from 30-71 minutes). Interviews were conducted remotely over Zoom and all participants agreed to our discussion being recorded. P13 preferred to conduct the interview over email instead of video call. Audio files were initially transcribed via Temi.com, a speech-to-text transcription tool. These automated transcripts were then manually corrected by the first author.  

Interviews followed a structured format with five segments: participant background, general experiences as a transgender individual in open source, the impact of their transition on OSS involvement, positive and negative OSS interactions, and attitudes toward OSS communities (see Appendix B for Interview Protocol). Initial questions focused on participant background and OSS career trajectory. We then asked participants to share their transition experiences and identity disclosure. Next, we asked about positive and negative experiences in OSS (which may or may not have been related to gender-identity). Finally, participants discussed discrimination in OSS towards trans contributors and recommendations they had for OSS communities broadly. At the end of each interview, we asked open-ended demographic questions where participants could self-identify. 

\subsection{Analysis}
We aimed to identify the participation challenges and benefits experienced by trans contributors and any impact of their gender identities and transitions had on their OSS career trajectories. In order to identify the influence of gender on OSS participation, we employed a Straussian grounded theory approach to analyze the data, thereby constructing theory based on patterns in the experiences of our participant and common emergent themes~\cite{strauss_grounded_1994}. Grounded theory is a methodology that has been used in the OSS literature previously \cite{steinmacher_social_2015, frluckaj_gender_2022, marlow_impression_2013}. We performed three rounds of coding as prescribed by Strauss: open coding, axial coding, and finally selective coding. 

\subsubsection{Open Coding}
The first author began open coding on five initial interview transcripts, focusing on participant accounts of the challenges and experiences as trans contributors in OSS. Themes were identified related to participant backgrounds, gender identity experiences, online representation, and OSS community recommendations. Memos written after each interview were incorporated, generating an initial set of codes to be expanded on. 
The first author compared subsequent interviews with the synthesis of newly generated and existing codes. Two authors then joined the coding process; three authors independently coded an additional two transcripts, generating codes for positive and negative participation experiences, instances of decreased or increased participation levels, and notable experiences for our participants. The three authors convened to discuss these transcripts, and the first two authors met more regularly to further develop codes through constant comparison. Over three coding and discussion sessions, these authors developed a preliminary set of 59 open codes from their independently coded transcripts. Open coding extended to the remaining interviews, expanding the set of codes. Interviews continued until no new themes emerged and theoretical saturation was reached. 

\subsubsection{Axial Coding}
We next performed axial coding on our full set of open codes to draw out shared experiences from our participants. The first author independently re-coded five transcripts using this new set of codes and then convened with the second author to discuss code application and disagreement. We used affinity diagramming to consider relationships among open codes to develop a set of initial higher level categories. We compared, contrasted, and consolidated the existing codes under emerging categories. Through an additional two rounds of code application and discussion, the codes were refined to 4 categories comprised of 25 codes and 168 subcodes.

\subsubsection{Selective Coding}
To understand the experiences of trans contributors in OSS participation, the first author employed selective coding focusing on participants’ OSS involvement. Codes from our codebook were then re-applied to consider fluctuations in participant involvement. The fourth author joined the project discussions at this stage, and subsequent analysis of emergent themes revealed patterns of behavior and experiences amongst our participants associated with increasing, decreasing or sustaining participation. We then referred back to the literature and extended our scope to more accurately understand new themes from the data.
We returned to the interview transcripts to ensure our resulting themes accurately reflected participants' experiences and language by searching for select keywords (e.g. ``privacy") and answers to specific questions (e.g. ``What do you think OSS communities could do better?"). We conducted member-checking by emailing all participants a copy of an article draft and asking for their reflections and reactions to the paper. We received positive feedback from 3 participants and no answer from the others. 

\subsubsection{Methodological Shift and Author Positionalities} Our analytical framing shifted during our return to the literature and latter stages of our analysis. At the premise of this study, we aimed to surface the ``unique" experiences of trans contributors in OSS. In doing so, we had unnecessarily Othered our participants in several categories. Through constant comparison, our research process changed to reconsidering our existing analyses and presentation of the results. We shifted to also include the similarities between the experiences of our participants and experiences exhibited in the literature. 
Finally, understanding the social locations of the authors is relevant to consider our lens on the data and our analyses. All of the authors are white, based in North America, and have studied open source software and participation. 
Author 1 is a queer first-gen Balkan graduate student who participates in their local queer community.  
Author 2 is a straight cis-man of middle age. He has participated in faculty governance LGBTQIA+ bodies as an ally. 
Author 3 is an US Arab-American cis-woman. She is an LGBTQIA+ ally.
Author 4 is a queer, trans first-generation college graduate and has been active in queer and trans open source communities and queer/trans activism for nearly two decades. 
Conducting this research was an example of allyship and solidarity through scholarship, however, it was not without missteps. Humility, patience, and openness were needed from all authors in order to learn from one another and move the research forward in a more inclusive and mindful way. 

\section{Limitations}
Like all research, our methods have limitations affecting the generalizability of our findings. By employing a qualitative approach, we relied on participants' verbal accounts rather than activity logs to understand their overall participation experiences. This reliance on self-reported data introduces potential biases and imperfect recall. Moreover, our sample size comprised of only 21 participants, reflecting the relatively small population of trans contributors in OSS. 
		
Our sample was not globally representative. Our recruitment email was in English and consequently, likely excluded participants who did not have working fluency in English, reducing the potential global reach of the study. The majority of our participants were white. We would have liked to include more racially diverse participants as the experiences of white trans people can be very different from those of POC trans people~\cite{pinter_entering_2021, noauthor_2015_2015}. Black and Latinx contributors have especially low involvement numbers in OSS~\cite{nadri_relationship_2022}. Unfortunately, given the small size of the participant pool, we were unable to specifically recruit for POC trans contributors. 

As discussed above, we recruited 13 of our 21 participants using GitHub's search function to identify potential participants, thus skewing our sample towards trans contributors who felt comfortable disclosing their identities publicly. Only 6 of our participants were not publicly out at the time of the interview. Furthermore, the researcher who conducted interviews is not trans. This also may have influenced how comfortable, connected, and understood participants felt, thereby impacting the quality of the interview data. 

Our participants were predominantly programmers who had all contributed technical contributions to OSS projects. We assessed participant activity based on their GitHub contribution graphs. Due to platform affordances, many non-technical contributions are made invisible. Thus we could not readily discover people who primarily contributed through design, documentation, or community management, which are other valid forms of contribution.

\section{Results}
While coding and discussing our results, we referred back to academic literature to compare emergent themes with existing understandings. During this process, we reflected on our positionality and membership in comparison to our participants; only one of our authors is trans. In grappling with the tension of membership ~\cite{liang_embracing_2021}, we transitioned from solely highlighting the differences between our participant experiences to those already documented in the OSS literature to also underscoring similarities and amplifying the messages that our contributors wanted to share. Thus, our results resist the lack of representation of trans experiences in the OSS literature while interweaving those experiences into the existing literature. In this section, we address our research questions and highlight experiences significant to our participants, including those involving the management and protection of their online identities while contributing to OSS. 

\subsection{Lessons from Trans Experiences in OSS}
\begin{mybox}
\textbf{RQ1: What did trans participants highlight about their experiences in OSS?} Participants discussed screening projects for perceived inclusivity before contributing and some experienced hidden microaggressions. Almost all participants advocated for both the inclusion \textit{and} enforcement of a Code of Conduct in OSS projects. On a positive note, our participants discussed OSS as a venue for their technical passions as well as a place to connect with other trans contributors. 

\end{mybox} 

\subsubsection{Screening for Project Inclusiveness and Codes of Conduct}

Our participants worked to “suss out” (P8) and inspect communities before formally deciding to join. Participants investigated the community’s language in issues, discussions, and other public forums. P1 described a project maintainer whose behavior within the project was not problematic, but upon digging deeper into the maintainer’s blog, found concerning behavior, deterring her from joining. Instead, P1 later moved to another project where the main project maintainer was openly non-binary, a signal that the project would likely be inclusive towards P1. A sense of inclusivity was integral for participants’ initial motivations in joining projects, as well as for their long-term retention.

P14 described multiple instances of it interacting with OSS projects at different extremes of inclusivity, from extremely trans-friendly communities to those filled with “literal Nazis.” P14 also described another method of screening projects for potential toxicity through the presence of a code of conduct (CoC) in the project. A CoC is a governance document that establishes the values and expectations of a particular community ~\cite{tourani_code_2017}. CoCs are designed to protect a community's members from harassment and attack, giving them a sense of security and belonging within the community~\cite{li_code_2020} and “signals a welcoming community”~\cite{qiu_signals_2019}. 15 of our 21 participants, including P14, stressed the importance of enforcement as it signals to trans and other under-represented contributors that they are supported by the broader community from harassment.

\begin{quote}
    \emph{Free software communities for me have been a coin flip of they're one of the kindest, most trans-friendly communities I have seen on the internet or they're chock full of literal, openly admitting, Nazis...Basically, if they've got a code of conduct, they're probably trans friendly. If they don't, they're probably bigots <laugh>. That’s a very rough rule of thumb and there's gonna be more nuance, but at a first glance, looking to see if they've got a code of conduct is a very good indicator of how trans-friendly the community will be.} \textit{(P14)} 
\end{quote}

P1, P2, P4, P6, P19 suggested that if a project is not committed to its enforcement, the presence of a CoC may lull contributors into a false sense of safety. Therefore, said our participants, projects should not include a CoC unless they are prepared to enforce it against their “best” contributor with appropriate consequences.

\subsubsection{Hidden Microaggressions}
Contrary to our initial expectations, participants in our study encountered minimal direct attacks based on gender identity. While broader online experiences included instances of transphobia, negative encounters in OSS predominantly manifested as what participants termed ``microaggressions" in contrast to direct, intention aggression. Participants often sensed subtle negativity related to their identity; the subtlety made it challenging to discern whether it was intentional or coincidental. This ambiguity led many to remain in spaces where they felt unwelcome, as they lacked concrete evidence to support their suspicions. P5 discussed how subtle microaggressions were more frustrating than targeted attacks as their opaqueness left room for debate. P5 attributed the constant uphill battle against discrimination and bigotry to one of the reasons that many [trans] people drop out of OSS. 
\begin{quote}
    \emph{The worst forms of discrimination are the subtle ones, because those are the ones that people fight you against. “That's not discriminatory. You just can't take a joke.” No, you just don't understand microaggressions. The big, loud mouths that are spewing hate and vitriol, those people are easy to deal with…those people go away pretty quickly because they get shut down and taken care of. It's the subtle microaggressions in day-to-day life and open source that really are what caused people to leave open source.} \textit{(P5)} 
\end{quote}
Our participants consistently dealt with (both online and offline) microaggressions which became more salient with each subsequent interaction. P8 described an instance where she promoted a very similar contribution as a (presumed cisgender) man, but her contribution was ignored while the man’s received community feedback and messages of interest. Repeatedly, our participants fought for their place in projects and against prejudices, subtle or not, which inevitably led to burnout. Another participant mentioned that while explicit transphobia might be encountered on more social media sites (e.g., Facebook and Twitter), those were less common on GitHub, which was more ``professional." Nonetheless, the more subtle forms of discrimination were found on GitHub. Our participants suggested these could be harder to deal with; the lack of insight into people’s social lives made it harder to contextualize their behaviors. 

\subsubsection{``You don't have to explain yourself over and over again''---Finding Community through OSS}

Despite the negativity that our participants encountered in OSS, they still had many positive experiences by finding other trans contributors to connect with and by receiving the support of an inclusive community. Connecting with other trans contributors in OSS removed a perceived barrier to entry for our participants: having to explain or justify their gender identities to projects with predominantly cisgender contributors. Instead they could collaborate with other trans contributors who held a more shared understanding of gender identity related experiences. P18 participated in several OSS projects but her most embedded involvement was with a project with other queer people. This meant she did not have to explain herself and could connect over shared experiences, which made joining and contributing to the project less stressful.

\begin{quote}
    \emph{I gradually moved into different communities...it's less stressful to hang out with other queer people...Obviously you talk about your experience with other queer people…you don't have to explain yourself over and over again. There are also...people who are not queer who are well informed, but they usually are more or less the minority.} \textit{(P18)} 
\end{quote}

Almost all of our participants mentioned the positive impact that an inclusive community had on their OSS experience, their personal lives, and careers. Many participants met with fellow trans contributors offline or in a distinct chat reserved for gender minorities due to an increased sense of kinship. These collaborations blossomed into networks, with trans contributors connecting one another, creating exciting opportunities for future work, job opportunities, and support. 

\subsubsection{Trans Co-visibility and Iconography}
\begin{figure}
  \centering
  \begin{minipage}[b]{0.48\textwidth}
    \includegraphics[width=\textwidth]{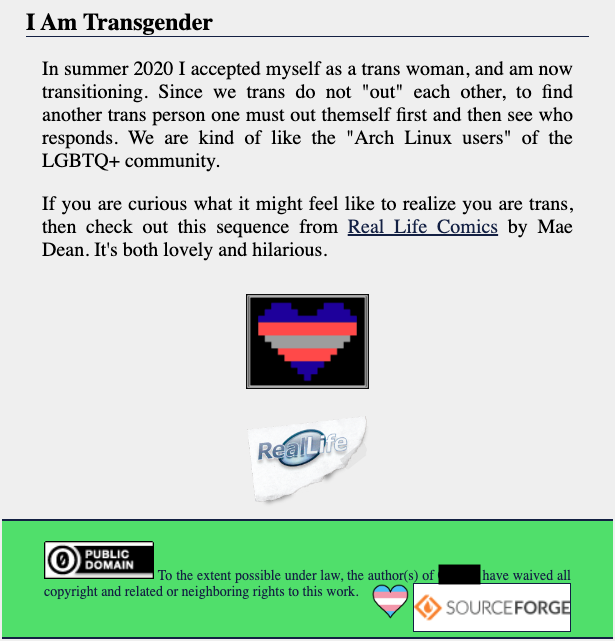}
    \caption{P3's project website include trans ``easter eggs'' --- trans iconography and a link to the comic that helped crack her egg.}
    \label{fig:1}
  \end{minipage}
  \hfill
  \begin{minipage}[b]{0.49\textwidth}
    \includegraphics[width=\textwidth]{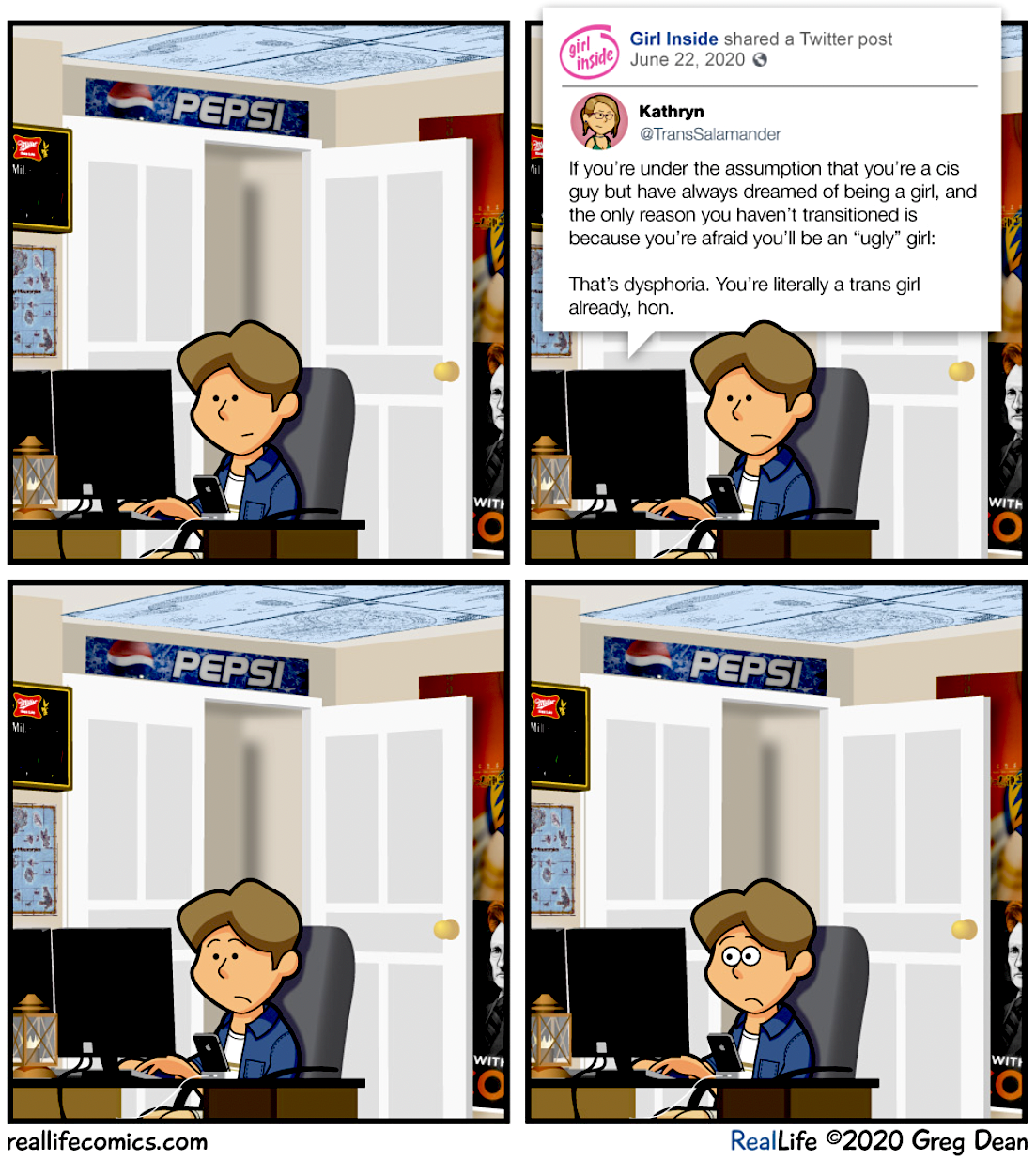}
    \caption{\textit{RealLife Comics} excerpt, cited by P3, where the comic author shares the story about their own realization and transition. Permission for use granted by author.}
    \label{fig:2}
  \end{minipage}
\end{figure}

Our participants were cognisant of the risks associated with openly displaying their transness in OSS. A few participants chose to hide their identities due to personal safety concerns, while others felt the potential benefits to other trans people in OSS outweighed the risks. Those who chose to risk their personal safety, like P4, felt that being visible was about something bigger than her, that it could be an inspiration to people struggling and signal that “they’re not alone.” Participants noted that upon seeing a trans or pride flag in a co-contributor’s GitHub bio, they immediately felt more comfortable engaging with them. On P3's project website, she included trans ``easter eggs'' (see Fig \ref{fig:1}) in the form of trans hearts, her own story, and a reference to a comic that helped her realize she was trans, shown in Fig \ref{fig:2}, thus signaling that the project was a safe space for trans contributors. The use of trans iconography and inclusion of their pronouns in our participants’ projects helped them share their identities, leading to many participants reaching out and finding new connections in OSS.

Several participants suggested that since many trans people are marginalized in the offline world, they turn to the online world for solace and comfort. Online, they found important resources and connections with other trans people that are otherwise unavailable to them in everyday life, leading to a growing presence and support network online. 

\subsection{Managing and Protecting Identity in OSS}

\begin{mybox}
\textbf{RQ2: How did our participants manage their online identities in OSS?} While transitioning, participants decreased or temporarily discontinued their OSS participation; post-transition, participants became interested in different OSS projects and communities. Most participants managed their online representation based on perceived audiences, which included grappling with their deadnames and contribution records associated with it. Finally, some trans contributors performed labor (usually in the form of moderation work) to maintain project inclusivity, which also led to burnout. 

\end{mybox} 

\subsubsection{Shifting Participation, Interests, and Communities}
All our participants answered our questions about transitioning. In our text below we use language like \textit{pre-} and \textit{post-transition}, while acknowledging that many queer theorists and people don't conceptualize transition with an endpoint~\cite{seeber_beyond_2015}. Perhaps influenced by our questions, participants used this language and did not flag it as problematic in member checking. As we coded and memoed these discussions and considered them in relation to our participants' OSS involvements, we came to understand the significance of transitioning in their lives---almost half of our participants stepped away from OSS while transitioning or became involved only afterwards. Several reported a general withdrawal from their previous activities and communities so that they could focus on themselves. 

Most of P8’s connections began post-transition, as she was relatively inactive in OSS before then: \emph{``I didn't have too many projects before I came out ... 98\% of people that know me, know me afterwards"} \textit{(P8)}. After her transition, P21 began to identify herself as part of a marginalized group and viewed the CoC, and how communities discussed it, differently. While the presence of a CoC had always been important to her, now it “really hit home” and became of a “first order importance” for her.

Several participants found their worlds were influenced in ways they described as surprising as they transitioned post-transition, including their OSS work. For instance, pre-transition, P3 focused on maintaining her mature projects, while post-transition, she opted to contribute less to work associated with her deadname and instead engaged in a mentoring-like role by contributing to younger projects.

\begin{quote}
    \emph{I don't put a lot of code out there right now, but I do tend to try to help existing projects, almost more like mentoring...This is definitely a trans thing. When I realized I was trans, a ton of my open-source time is there because I was closeted to myself, I was in the egg, and I was just kind of passing time. And so that is a hobby that consumed a ton of time when I thought I was a boy, that now I know I'm not, I need that time for other things. I guess my interests changed sharply when I came out to myself.} \textit{(P3)} 
\end{quote}

During transition, many of our participants cut ties with the “toxic” people in their lives. Past connections were less emphasized while other interests, such as connections to other trans contributors and finding a sense of belonging in OSS communities, became more important post-transition. 

\subsubsection{Identity Separation and Avoiding the In-Laws}
Participants felt a disconnect between their offline and online representations post-transition. While our participants were at different stages of the ‘coming out’ process, all mentioned a hesitancy to do so in their offline lives for fear of their personal safety, and occasionally exhibiting a defensive mindset when interacting with new acquaintances. P11 describes their two separate personas online and offline, as they did not feel safe or comfortable presenting as a non-cis man: \emph{``I have stayed more on the periphery of open source...I have been well aware of people being hostile to anyone who is not a cis man...I have not personally had many bad experiences in open-source, but it's kind of because of this defensive strategy"} \textit{(P11)}.

While still remaining relatively anonymous online, (e.g., using pseudonyms/nicknames on profiles), participants felt much more comfortable disclosing their emotions, opinions, and experiences to online friends. Multiple participants came out to their online friends first, rather than offline friends and family. P21 was very selective as to where she updated her profile as she had not come out to her extended family yet, and those changes may have repercussions in her offline personal life. Despite P21's use of GitHub as a professional site, the openness of it enabled surveillance by her in-laws, and so she avoided updating her profile there. 

\begin{quote}
    \emph{I'd like to change it as soon as possible, but...there are a few things that have to happen first. Because I'm sure my parents-in-law keep an eye on my GitHub account.} \textit{(P21)}
\end{quote}

Hesitancy for trans people to truly reflect themselves in more open online spaces and in offline spaces directly resulted from fear of being prematurely outed, personal attacks on their physical person, or even fear of their careers suffering. 

\subsubsection{Deadnames in Digital Archives}
Almost all of our participants preferred using a name other than their deadname – P7 was the only participant who kept their birth name as it was already gender neutral. While our remaining participants had names they preferred to be addressed by, not everyone had changed their profiles or accounts to reflect this change. This was for a number of reasons, i.e. they had not come out to their families yet and wanted to keep their gender identity a secret for now or did not want to lose past work associated with their deadname. Some participants viewed the change as a fresh start and did not necessarily lament the loss of work tied to their deadname. P4 was particularly attached to a handle she had been using for many years, but felt that it no longer represented her so she was happy to discontinue it. 

\begin{quote}
    \emph{I had been using the same handle since the early 90’s. So it was kind of a big deal for me to leave that behind, but it felt attached to a person that I wasn't anymore.} \textit{(P4)}
\end{quote}

For the majority of participants, however, having their immutable deadname connected to their work made navigating their new identities treacherous. P9 spoke of the difficulty in changing their name in GitHub and in git, \textit{``Some people change their names and then they have to change all their git history or they just lose their commits"} (P9). If they are unable to change all references to their deadname, they may risk losing all their past contribution history. 

P16’s own difficult experience reclaiming work associated with their deadname left them advising future programmers to choose names that are not connected to their real name as that username would remain unaffected by the transition --- \emph{``My advice to any developers out there is if you're planning on changing your gender later on...don't use your real name in the git records"} (P16).

\subsubsection{Maintaining Project Inclusivity}
Many of our participants had taken on, implicitly or explicitly, the role of moderator to protect themselves and their communities from the bigotry and transphobia witnessed in other contexts. P14 requested to be placed on the moderation team of its project \emph{``to make sure we don't turn into a transphobic hellhole."} At the time of our interviews, however, most of our participants were exhausted with this role or had completely ceased their moderation efforts due to burnout. These negative experiences were largely attributed to two factors: the deep acting and emotional labor~\cite{hochschild_managed_2012} of moderation work and the inadequate support received from the broader project community. Moderation is not a straightforward or “easy” form of OSS contribution as tasks can vary from corralling a dissatisfied user to identity-based arguments with bigots.

\section{Discussion}

As we began to connect the empirical evidence and results from our interviews with trans contributors with the existing OSS literature, we saw many parallels in the experiences and grievances expressed by our participants. This was contrary to our initial intention of documenting the ``unique" experiences of trans OSS contributors. Instead, we learn from our participants' experiences to extrapolate and theorize for OSS and open collaboration more broadly. In this section we highlight two paradoxes: openness and display and openness and governability. We discuss how they interact, theorizing processes that could lead to the results presented above.

\subsection{The Paradox of Openness and Display}
Open collaboration depends on visible records. Indeed one definition of open collaboration is ``actionable transparency" --- the ability to see what is going on, as well as the ability to jump in and act)~\cite{colfer_mirroring_2016}. OSS facilitates broad and decentralized participation, simultaneously creating visibility and scrutiny. For participants, this meant not only that one’s own behaviors are visible, but also that participation required witnessing the behaviors of others (even if one was not directly involved in that discussion). Further, those behaviors are recorded and are expected to remain visible for the indefinite future. This display of behavior causes contributors to manage their own behavior based on perceived audience(s), as well as reviewing and responding to the display of others in OSS projects. Our data makes it clear that the consequences of this are paradoxical. We first summarize, then discuss both aspects in more detail below.

On one hand, participants spoke of the autonomy over their own display that online participation enabled, echoing some of the earliest hopes for online participation enhancing inclusion~\cite{malinen_understanding_2015, bargh_internet_2004}. This included being able to control presentation of gender in text-based communications, but also less discussed elements such as increasing autonomy over where one can live, enabling participants to live in less trans-antagonistic jurisdictions.

On the other hand, this heightened level of control over display imposed additional record keeping effort and multiple audience management~\cite{bernstein_quantifying_2013, kairam_talking_2012}. Participants had to hide their profiles from select audiences while also wanting to connect with other trans contributors. Participants felt compelled to maintain inclusive records and signals of their own projects---this meant uneven distribution of moderation labor over the community records.

\subsubsection{Disembodiment in OSS Work}
Participants valued the disembodied experience of contributing online, citing benefits such as flexibility and avoidance of unwanted in-person interactions, each features of remote work discussed in the literature~\cite{smite_work--home_2023, bartman_promote_2021}. Remote work, known for offering alternatives and flexibility, can also mitigate discrimination experiences~\cite{hickox_remote_2020}, particularly for trans contributors, as it potentially allows for controlled identity disclosure, high-impact technical work, and the autonomy to disengage and re-engage~\cite{ford_how_2019}. Participants spoke of temporarily circumventing the constraints and burdens of the offline world; they did not worry about how they dressed nor how people would perceive them (or misgender them), they were able to choose their physical location (e.g. P1 didn’t want to contribute from North Carolina due to legislation against gender-neutral bathrooms), and were able to separate their online and offline lives. The disembodiment afforded by OSS work, and remote work more broadly, helped reduce the visibility  of historical characteristics for discrimination (e.g. gender, race, age, etc.) and provided participants autonomy over their online displays. 

During the interview phase, several participants preferred to speak with their cameras off, or with the camera on but the lights off – obscuring their face, or with a VR avatar in their stead. Remote online work provides an opportunity to exist, connect, and contribute invisibly. That our participants highlighted (and used) ways that online communication enables autonomy in identity presentation hearkens back to early ideas in CSCW and Computer Mediated Communication. Indeed, Asenbaum reminds us that the trope of invisibility of stigma and its contradictory effects in online spaces extends back a long way~\cite{asenbaum_theoretical_2017}, with Hiltz and Turoff writing in 1978~\cite{ hiltz_network_1993}:

 \begin{quote}
     \emph{General appearance, such as height, weight, and other culturally determined aspects of “attractiveness” and the clothes, makeup, jewelry, and other props used by persons to present themselves to others, provide an important filtering context for face-to-face communication. So do the visibly apparent cues that are provided by sex, age, and race and by visually apparent physical handicaps. In general, those aspects of self that are devalued by a culture – such as being black, female, old, “ugly”, or disabled – have the effect of acting as a general stigma...[Through online anonymity, however,] it is the content of the communication that can be focused on, without any irrelevant status cues distorting the reception of the information, especially if anonymity makes even the sex of the contributor unknown.} (Hiltz and Turoff 1978, p.78, 91)
 \end{quote}

Despite much criticism of these concepts~\cite{boellstorff_coming_2015, daniels_rethinking_2013, blackwell_harassment_2019}, it was inescapable in our interviews that participants valued aspects of these ideas, yet their endorsements were always tempered and partial. Our analysis finds this tempering to be three-fold: participants had to work hard to manage the records associated with them, felt that they were limited in their ability to contribute authentically, and had to manage the records of the community around them. By regarding these issues in the context of our participant’s voices, we further theorize on the implications of openness, visibility, and governance in OSS and beyond. 

\subsubsection{Managing Individual Displays}
First, participants spoke of difficulties with autonomous presentation, stemming from the record-keeping functions of the online infrastructure. Only recently did GitHub offer the possibility of changing identifiers in archives. Yet even that is insufficient, since git (a separate system from GitHub's database backed website) contains identifiers on each commit, e.g. a name and email address. These are harder to change even within a single repository (requiring editing git history) and since git-based collaboration results in many distributed copies of repositories through forks and clones there are no automatic ways to change identifiers in repositories controlled by others.

Participants deliberated the transparency of their GitHub profiles regarding gender identity, acknowledging both benefits and drawbacks. Some used clear signifiers of their gender identities indicated their pride (P2, P3, P8, and P15) and that they were open to trans collaborators (P3, P4, P6) and to being mentors for hidden trans contributors (P4, P10, P17). However, concerns about trolls and harassers prompted others to prefer hidden profiles for reduced risk. Yet, hiding also limited connections with fellow trans OSS contributors. P12 suggested using encoded language to signal a trans identity to fellow contributors while evading detection from outsiders. This practice echos the Victorian ``language of flowers''~\cite{seaton_language_2012}, e.g. green carnations signaling a queer identity~\cite{beckson_oscar_2000}. 

Online platforms like GitHub obscure information crucial for sense-making, such as body language, leading to difficulties conveying nuance and complexity. Social bubbles and echo chambers thrive, exacerbating the issue~\cite{eady_how_2019}. GitHub's professional setting encourages hidden microaggressions, as contributors struggle to contextualize behavior due to limited insight into others' lives, often projecting gender onto participants. While projected misgendering was common, some evidence suggests that these projections and mental models are more malleable in online settings than they would be in offline settings~\cite{qin_is_2021}. P16, who collaborated with developers both in their physical place of work and online, noted that post-transition, their remote collaborators adjusted and began using their correct name and pronouns much more quickly and effectively than their offline collaborators.

\subsubsection{Contributing Authentically}
Participants spoke of wanting to present their authentic selves and to participate in projects with like-minded participants where they did not feel they had to rigorously control their gender presentation; they felt that they valued autonomy but spoke about its necessity as unjust and constraining. The openness of OSS participation led to feelings of elevated scrutiny and privacy risks for our participants, contributing to context collapse~\cite{marwick_i_2011}.

Open online participation carries risks for marginalized populations~\cite{barakat_community_2022, benjamin_race_2019}. Many of our participants perceived privacy concerns while contributing openly, adjusted in ways that made their authentic contributions and gender expression more difficult~\cite{thomas_diversity_2020}. Increased visibility of trans people has contributed to growing violence against them~\cite{robinson_lavender_2020}. Despite well-intended efforts to increase visibility of trans people to promote social change, Beauchamp (2019) reminds us that visibility is not a panacea, but rather as Foucault remarked, a trap~\cite{beauchamp_going_2019, foucault_discipline_2012}.

Participants related how hidden profiles provide respite from online harassment but also require constant vigilance, hindering authentic and sustainable OSS participation. While hidden profiles may enhance perceived security, they disconnect contributors from their OSS involvement, impeding learning, networking, and career opportunities. OSS has long been lauded as a site of situated learning and legitimate peripheral participation~\cite{lave_situated_1991, nafus_patches_2012}, yet this reputation is problematized by the experiences of our participants because the communities of practice literature has not addressed the impact of hidden profiles on participation.

\subsubsection{Managing Communal Displays}
Finally, participants' stories emphasized that despite autonomy in presentation, they still had to witness others' behaviors, extending beyond responses to their own presentations; participants needed to witness publicly visible discussions, which could be problematic and harmful to community reputation, between others in the project to fully participate. Many of our participants had to “suss out” (P8) opaque communities to determine how welcome or safe they would be. It was difficult to assess whether a community condones toxicity~\cite{ragkhitwetsagul_toxic_2021, sarker_automated_2023} or prioritizes inclusive behavior prior to contributing; participants investigated the project community’s language in issues, discussions, and other public forums, and noted the presence of a community CoC. 

Participants often performed reconnaissance work to select inclusive projects to contribute to and frequently assumed moderation roles to ensure community inclusivity, addressing transphobic or bigoted behavior by enforcing CoC guidelines, warning community members of potential infringements and thereby creating shared learning and culture ~\cite{cullen_practicing_2022}, and by removing infringers. In P14’s case, it and its co-moderators deserted their project when the project leader refused to condemn transphobic behavior, contributing to P14's eventual burnout. Moderation work is often performed by under-represented groups~\cite{rodriguez_addressing_2015, williamson_minority_2021}, and women and trans contributors frequently perform community-oriented labor in OSS \cite{trinkenreich_womens_2022, frluckaj_gender_2022}. The emotional labor of moderator and community maintenance is also unevenly distributed amongst community members, contributing to burnout \cite{riedl_downsides_2020, schopke-gonzalez_why_2022}. The openness of OSS disproportionately burdens under-represented contributors as they labor to not only manage their own visibility, but that of their communities as well. 

In summary, the visibility of records axiomatic to participating in OSS development allows contributors to exert significant control over their displays but also requires an immense amount of labor in managing one's own records and those of one's communities. While our participants benefited from the flexible and remote nature of OSS work, they had to vigilantly self-monitor their profiles, impacting authentic and sustainable participation. Furthermore, as it was in our participants best interest to maintain inclusive community displays, much of this management and moderation work was unevenly distributed to our participants within the projects they contributed to. This ties into the next paradox, that of openness and governability in OSS.

\subsection{The Paradox of Openness and Governability}
A key aspect of openness is the ability for anyone to jump in and act, without prior approval. Potential participants can undertake work relevant to them and then offer it to others, all without pre-approval processes (e.g. passing through an organizational hiring process or passing required organization process and value training). Organizationally, the more open a project, the more it exists outside what have been called institutional containers~\cite{winter_beyond_2014}. In OSS, unlike social media sites or online games~\cite{schoenebeck_drawing_2021,schluger_proactive_2022,ma_esports_2022,blackwell_harassment_2019}, the platform provider has not generally been seen as the primary provider of governance---rather individual projects have.

The surprisingly capable infrastructure of collaboration offered by platforms such as GitHub means that complex work can occur prior to the existence of surrounding institutional structures (no fundraising or investment is needed to purchase the technology needed to support work). The broad decentralization and openness of OSS facilitate open collaboration that transcends institutional boundaries yet simultaneously remove resources for accountability and governance, meaning that successful projects have to build their own approaches~\cite{shah_motivation_2006,omahony_governance_2007,de_noni_evolution_2013}. 

We describe this as a paradox of openness and governability: openness is attained by reducing institutional structures, but these structures also constrain behaviors. As the shared culture and guidelines of organizations vanish, so do their governing bodies like HR or compliance departments. There often is no ``skip-level manager'' to report behaviors to~\cite{li_code_2020}. While openness allows anyone to participate, it also leaves individuals with fewer tools to influence others' behaviors; our participants explained how this burden fell disproportionately on them than on majority participants.

The key resource our participants were left with was discourse: entering the fray of discussion to argue for one's perspective. Discourse itself requires resources. CoCs and the processes of moderation provide grounds to which participants can point to for arguing the unacceptability of certain behaviors. CoCs create common ground, a key resource in discourse~\cite{clark_using_1996, koschmann_trouble_2016}. Our results make it clear that our participants felt that cis participants were unlikely to engage with and moderate misbehaviors, placing the burden of correcting behavior on those we interviewed. As behaviors were highly visible, participants felt obligated to identify infringements and often had to persuade others that the infringement was indeed harmful; they spoke of the difficulty of repeatedly being required to try to build common ground. Their moderation efforts relayed a commitment to inclusivity, performed without institutional resources to support them. Given the intensity of this labor and connection to their personal identity, marginalized contributors are more vulnerable to burnout and discontinuing OSS participation.
  
To feel more supported, many of our participants contributed to and preferred being in OSS projects with other trans contributors. Not only did their shared identities help reduce the barrier to entry for the project, but it also helped facilitate community and solidarity. Trans contributors could vent to and rely on one another in instances of perceived microaggressions, as many had experienced bigotry in offline and online spaces. This resulted in what our participants described as expansive and deeply connected networks of trans contributors where they could help reduce scrutiny of their gender identities, support one another, and recommend inclusive projects, akin to whisper networks for women~\cite{johnson_purpose_2023}. As openness, supported by capable work platforms, rises, participants lose the governance resources built within traditional organizations; discourse dominates and finding participants with pre-existing common ground would be especially valuable.

In summary, we present two paradoxes associated with openness and discuss how these paradoxes interact. The paradox of openness and display is that open participation requires record creation and thus witnessing and archiving the public behavior of others. This tempers benefits from the experience of autonomy of self-presentation afforded by textual remote communication. The risks of display, and the impact of witnessing problematic behaviors, falls unequally on the shoulders of non-cis participants for whom display has higher risks. Here the paradox of openness and governability interacts: as openness increases, traditional resources for addressing injustice in organizations fall away. This leaves participants to rebuild these resources, drawing on CoCs, but never without the tiring and unequal work of jumping into discussions, attempting to build common ground, and to fight for understanding, empathy, and justice. Our participants embraced that challenge, but also spoke of frustration that it was needed, and sought projects in which more common ground already existed. 

\section{Recommendations}
Drawing from the experiences of our participants and our theorizations regarding the openness of OSS, we present recommendations to support more inclusive participation.

On openness and display, platforms should enhance support for managing individual and community displays and records. GitHub and other social networking sites should facilitate more privacy controls and customization for individual contributors (e.g. toggling the privacy/availability of records, add the ability for transitive changes in git edit history to clones and forks), should reduce the labor of underrepresented contributors witnessing and identifying misbehavior (e.g. algorithmic support in identifying microaggressions ~\cite{schluger_proactive_2022, hsieh_nip_2023}) and introduce support for collective accounts to foster community solidarity and sustained participation (e.g. P13 was part of a pluralqueer system which shared one account on most social media sites but had multiple distinct GitHub accounts). The Queer HCI literature could more deeply explore trans and queer open collaboration to determine how to better support queer contributors and their community bonds, as well as collaboration between queer and non-queer contributors.

Regarding openness and governability, OSS communities should improve moderation training and distribute moderation tasks across the community, rather than underrepresented contributors performing that labor with little help. Moderation is difficult and draining work; OSS projects should provide resistance training and attitude inoculation~\cite{fagnot_enhancing_2015} for their moderators. At the project level, moderation should be more evenly distributed (e.g. moderation labor acknowledged as a form of recognized “service” that contributors can rotate through and perform temporarily). At the broader environment level, aligning with Jhaver et al. and Zannettou, we advocate for enhanced policies on social networking sites like GitHub and Twitter to deter harassers and safeguard marginalized users more effectively~\cite{jhaver_evaluating_2021, zannettou_i_2021}.

Future research should investigate how openness in OSS effects privacy risk perception for underrepresented contributors and explore how contributors balance perceived benefits with risks. Future research should also explore case studies of online moderation, and what enables successful and sustainable moderation equitably shared across projects.  Research on communities of practice and situated learning should also address questions such as how can participation where participants do not feel safe presenting their authentic selves affects dynamics in communities of practice? Future research should investigate the impact of profile visibility on learning, legitimacy, and collaboration. Finally, further research is needed to provide inclusive direction for OSS projects and GitHub more broadly.

\section{Conclusion}
While there is still much to be achieved for inclusivity and equity within OSS, this qualitative study on trans OSS contributors connects their challenges and barriers faced in OSS to the experiences of contributors already known and expands Queer HCI literature to open collaboration. Our results enabled theorizing about two paradoxes in open collaboration: the paradox of openness and display, and the paradox of openness and governability. 
Research and practice needs to advance work to understand the implications of openness.  While there is great potential for equitable and diverse participation, perhaps beyond that in traditional organizations and face to face communities, there are persistent poor outcomes and experiences~\cite{dunbar-hester_hacking_2019}. In this paper we have brought forward trans experiences in OSS and contributed to theorizing ways in which openness presents paradoxes for diverse participation, centered around public records that display interactions and resources for governability. We hope that our theorization and the experiences we present will help OSS participants and communities as they work, lending support to a world where, in the words of a participant, OSS projects will \emph{``make it a rule to be accepting to different minorities and LGBTQ+ peoples and make it very clear that the people controlling the space will be supportive if it's ever necessary" (P10)}.

\begin{acks}
This work would not be possible without the cooperation, patience, and enthusiasm of our participants, and for this we thank them. We thank Alex Ahmed, Bogdan Vasilescu, and Lee Kravchenko for their comments and feedback on this study.
This material is based upon work supported by the Sloan
Foundation under Grant No. G-2020-14036 and the National Science Foundation under Grant No. 2107298. 
\end{acks}

\bibliographystyle{ACM-Reference-Format}
\bibliography{references}

\pagebreak
\section{Appendix}
\subsection{Appendix A - Participant Demographic Information}

\begin{table}[htbp]
\begin{tabular}{l|l|l|l|l|l}
\multicolumn{1}{c|}{\textbf{P\#}} & \multicolumn{1}{c|}{\textbf{\begin{tabular}[c]{@{}c@{}}Self-identified \\ Gender Identity \\ (Open Response)\end{tabular}}} & \multicolumn{1}{c|}{\textbf{Pronouns}} & \multicolumn{1}{c|}{\textbf{\begin{tabular}[c]{@{}c@{}}Publicly out\\ at time of \\ interview?\end{tabular}}} & \multicolumn{1}{c|}{\textbf{\begin{tabular}[c]{@{}c@{}}Racial/Ethnic \\ Identity (OR)\end{tabular}}} & \multicolumn{1}{c}{\textbf{\begin{tabular}[c]{@{}c@{}}National \\ Origin (OR)\end{tabular}}} \\ \hline
1 & Trans-feminine & she/they & Yes & \begin{tabular}[c]{@{}l@{}}Non-Hispanic \\ and White (NHW) \end{tabular} & USA \\ \hline
2 & Female & she/her & No & \begin{tabular}[c]{@{}l@{}}NHW\end{tabular} & USA \\ \hline
3 & Woman & she/her & Yes & \begin{tabular}[c]{@{}l@{}}NHW\end{tabular} & USA \\ \hline
4 & Female & she/her & No & \begin{tabular}[c]{@{}l@{}}NHW\end{tabular} & GER \\ \hline
5 & \begin{tabular}[c]{@{}l@{}}Non-binary trans \\ woman\end{tabular} & she/her & Yes & \begin{tabular}[c]{@{}l@{}}NHW\end{tabular} & USA \\ \hline
6 & Woman & she/her & No & \begin{tabular}[c]{@{}l@{}}NHW\end{tabular} & \begin{tabular}[c]{@{}l@{}}Occupied Caw \\ land in USA\end{tabular} \\ \hline
7 & Non-binary & they/them & No & Southeast Asian & AUS \\ \hline
8 & Trans woman & she/her & Yes & \begin{tabular}[c]{@{}l@{}}NHW\end{tabular} & USA \\ \hline
9 & Queer trans woman & they/them & Yes & African-American & USA \\ \hline
10 & Female & she/her & Yes & \begin{tabular}[c]{@{}l@{}}NHW\end{tabular} & USA \\ \hline
11 & Non-binary & NA & Yes & \begin{tabular}[c]{@{}l@{}}NHW\end{tabular} & NL \\ \hline
12 & \begin{tabular}[c]{@{}l@{}}Genderqueer and \\ genderfluid\end{tabular} & he/his & No & Asian & USA \\ \hline
13 & NA (email participant) & NA & No & NA & NA \\ \hline
14 & \begin{tabular}[c]{@{}l@{}}Non-binary trans \\ woman\end{tabular} & it/its & Yes & \begin{tabular}[c]{@{}l@{}}NHW\end{tabular} & USA \\ \hline
15 & Female & she/her & Yes & Undisclosed & USA \\ \hline
16 & Female & she/her & Yes & \begin{tabular}[c]{@{}l@{}}Hispanic and \\ White\end{tabular} & USA \\ \hline
17 & Undisclosed & she/her & Yes & \begin{tabular}[c]{@{}l@{}}Hispanic and \\ White\end{tabular} & USA \\ \hline
18 & Female & she/her & Yes & \begin{tabular}[c]{@{}l@{}}NHW\end{tabular} & GER \\ \hline
19 & Trans femme & she/they & Yes & \begin{tabular}[c]{@{}l@{}}NHW\end{tabular} & USA \\ \hline
20 & Non-binary & they/them & Yes & Indian-American & USA \\ \hline
21 & Trans woman & she/her & No & \begin{tabular}[c]{@{}l@{}}NHW\end{tabular} & UK

\end{tabular}
\caption{Participant Demographics - Gender identity, racial/ethnic identity, and national origin were open response questions for participants to self-identify.}

\label{tab:projects}
\end{table}
\pagebreak
\subsection{Appendix B - Interview Protocol}
\begin{enumerate}
    \item{Background}
    \begin{itemize}
        \item Tell me a bit about who you are and what you do.
        \item (If pronouns or gender identity unknown): How do you identify your gender?
        \item How did you first learn about open source software?
        \item Can you tell me about your early experiences in open source?
        \item How are you currently involved in open source? Which communities do you interact the most with and how?
    \end{itemize}
    \item{General Gendered Experience}
    \begin{itemize}
        \item Can you tell me what it's like to be a trans/non-binary person in open source?
        \item Has gender identity come up in your interactions with people in OSS?
        \item Can you briefly describe some gay, lesbian, bisexual, transgender, non-binary, or queer people who are important to you in your developer community?
        \item Are there any groups you are a part of that are related to being trans in open source or computing? 
    \end{itemize}
    \item{Transition Experience}
    \begin{itemize}
        \item Did you let people know about your gender? Who and how?
        \item Can you tell me how your connections in your developer community reacted to your experience? (If so, how? / If no, why not?)
        \item Did you make your change in gender identity publicly known somehow? How? Why or why not?
        \item How did you manage your online representation? (Did you ever create a different account for open source work to reflect your change in gender identity?)
        \item How did topics related to your gender identity come up in your interactions with people in OSS? 
    \end{itemize}
    \item{Positive and Negative Experiences}
    \begin{itemize}
        \item Can you tell me about any positive interactions you've had in open source?
        \item Can you tell me about any negative interactions you've had in open source?
    \end{itemize}
    \item{Attitude}
    \begin{itemize}
        \item What does the word "discrimination" mean to you?
        \item Have you ever experienced discrimination?
        \item Have you ever heard stories about an LGBTQ+ person being discriminated against within open source?
        \item How do you think OSS communities could do better?
        \item Is there anything else you think we should know about your involvement in open source or software development?
    \end{itemize}
   
\end{enumerate}

\end{document}